\documentclass{article}
\usepackage{spconf,amsmath,graphicx}

\usepackage{enumitem}
\setlist{nosep, leftmargin=14pt}

\usepackage{booktabs}
\usepackage{xcolor}
\usepackage{multirow}
\usepackage{makecell}
\usepackage{soul}
\usepackage[textsize=tiny]{todonotes}

\usepackage{caption}
\title{Domain generalization in fetal brain MRI segmentation \\with multi-reconstruction augmentation}
\name{\begin{tabular}{c}Priscille de Dumast$^{1,2}$, Meritxell Bach Cuadra$^{2,1}$\end{tabular}}

\address{$^{1}$ Department of Radiology, Lausanne University Hospital (CHUV) and \\ University of Lausanne (UNIL), Lausanne, Switzerland \\
$^{2}$ CIBM Center for Biomedical Imaging, Switzerland }

\begin{document}
%\ninept
%
\maketitle
\begin{abstract}
Quantitative analysis of \textit{in utero} human brain development is crucial for abnormal characterization.  
Magnetic resonance image (MRI) segmentation is therefore an asset for quantitative analysis. 
However, the development of automated segmentation methods is hampered by the scarce availability of fetal brain MRI annotated datasets and the limited variability within these cohorts. 
In this context, we propose to leverage the power of fetal brain MRI super-resolution (SR) reconstruction methods to generate multiple reconstructions of a single subject with different parameters, thus as an efficient tuning-free data augmentation strategy. Overall, the latter significantly improves the generalization of segmentation methods over SR pipelines. 
\end{abstract}
\begin{keywords}
Magnetic resonance imaging (MRI), Super-resolution (SR) reconstruction, Automated fetal brain tissue segmentation, Data augmentation, Domain adaptation
\end{keywords}
\vspace{-.2cm}
\section{Introduction}
\label{sec:introduction}
\vspace{-.2cm}
Human brain undergoes most significant changes \textit{in utero}. During the prenatal period, disruption of maturation processes may lead to abnormal development resulting in severe conditions such as congenital diseases, developmental delay or cognitive impairment later in life~\cite{egana-ugrinovic_differences_2013, volpe_perinatal_2001}. Magnetic resonance imaging (MRI) is a complementary imaging modality for prenatal diagnosis as it has proven its clinical value for the assessment of intracranial structures~\cite{garel_developpement_2000,gholipour_normative_2017, griffiths_change_2017}. Structural T2-weighted (T2w) MRI offers a good soft-tissue contrast to monitor the fetal brain growth and tissue maturation. Being very sensitive to the stochastic fetal motion, fast 2D acquisition schemes are used in order to minimize intra-slice motion~\cite{gholipour2014fetal}. The downside effect of such acquisition stands in the strong anisotropy of the resulting low-resolution (LR) images with remaining inter-slice motion.

In the last decades, super-resolution (SR) algorithms have offered the possibility to reconstruct a single high-resolution (HR) motion-free isotropic volume from a set of LR orthogonal acquisitions~\cite{kuklisova-murgasova_reconstruction_2012, ebner_automated_2020,tourbier_mialsrtk_2020}. Such SR reconstruction (SRR) methods leverage the redundancy in LR images to estimate 
% the fetal brain motion in a slice-to-volume registration approach. 
inter-slice inter-series motion.
Subsequently, a HR image is restored by solving an inverse problem in which a regularization function in considered with varying weight.
As previously discussed in~\cite{payette_efficient_2020}, different %intensity-based preprocessing and
reconstruction % regularization 
translates into a substantial variation in the reconstructed image appearance (Fig.~\ref{fig:results_regularization}).

Although SR-reconstructed volumes allow the analysis of 3D imaging biomarkers~\cite{kyriakopoulou_normative_2017,khawam_fetal_2021}, yet morphometric and volumetric measures remains restricted as it strongly relies on the MR image tissue segmentation. Manual annotations are time-consuming and prone to inter-rater reliability, hence, automatic segmentation methods are a key asset for further quantitative analysis~\cite{payette_automatic_2021,payette_fetal_2022}. 
However, automatic fetal brain MRI segmentation remains challenging as it is subject to many domain shifts, amongst which the SR reconstruction pipeline. Indeed, the many different intensity-based operations and regularization performed during SRR induces an inter-SR method data distribution shift that is for the segmentation method to overcome.

Recent works \cite{payette_efficient_2020, dedumast_syntheticMRI_2022} evidence the need for domain adaptation strategies %, such as transfer learning, 
to fit the SR domain gap in fetal brain MRI segmentation. While~\cite{payette_efficient_2020} uses noisy registration-based labelling between same-subject different-SR method volumes, \cite{dedumast_syntheticMRI_2022} takes advantage of synthetic fetal brain MRI to increase the training sample size in the SR target domain.
More generally, domain-adaptation strategies for MRI segmentation have been proposed~\cite{ouyang_causality_inspired_arxiv_2021, OrbesArteaga_augmentation_arxiv_2022}. Nevertheless, such methods are all data-driven, hence adjustment to the target task requires optimization efforts. Moreover, while these methods specifically aim to integrate non-linear intensity changes, these are not specifically addressing the real SR domain shift.

In this paper, we propose for the first time to exploit in-SR domain multi-reconstruction, i.e. reconstructions of the same brain obtained with different regularization weight, as an intensity-based data augmentation for intracranial tissue segmentation of fetal brain MRI.
In that manner, we hypothesize that the intensity variability of the training samples of a segmentation model is increased, without increasing the need for manual annotation.
We will show also the domain generalization power of our original multi-reconstruction approach on a pure out-of-SR domain dataset.
\begin{figure*}[h!]
    \centering
    \includegraphics[width=0.72\linewidth]{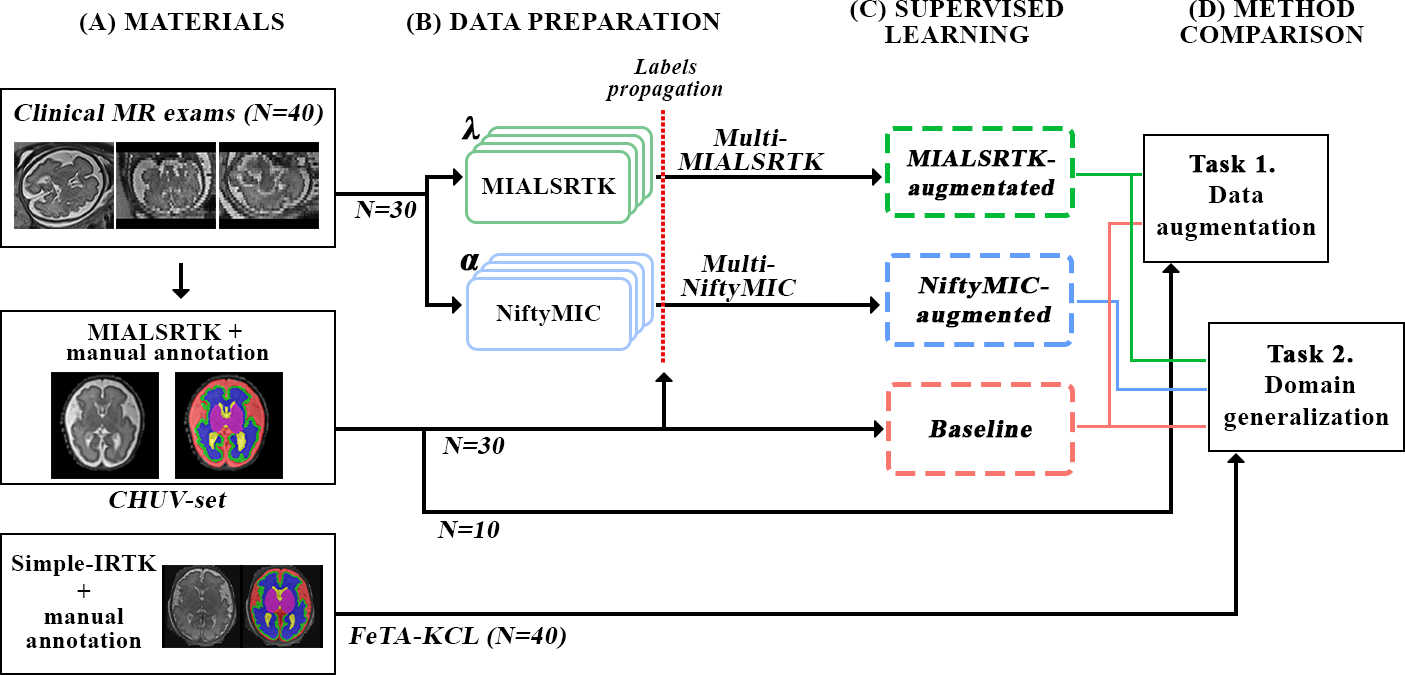}
    \caption{Overall framework. }
\label{fig:experiment_design}
\end{figure*}

\section{Materials}
\subsection{Clinical MR exams}
\label{ssec:clinical_exams}
\vspace{-.2cm}
Fourty (40) clinical fetal brain MR exams from 21 to 35 weeks of gestational age (GA) (mean $\pm$ standard deviation (SD): $28.4 \pm 4.2$) were retrospectively collected from our institution. Acquisitions were either performed at 1.5T ($N=37$) or 3T ($N=3$), respectively resulting to $1.125 mm$ and $0.5469 mm$ in-plane isotropic, and $3.3 mm $ and $3 mm$ through-plane resolution.

\vspace{-.2cm}
\subsection{In-SR domain dataset -- \textit{CHUV-set}}
\label{sssec:feta_chuv}
\vspace{-.2cm}
All clinical exams (Section~\ref{ssec:clinical_exams}) were SR-reconstructed into the subject's space through the MIALSRTK pipeline (with default regularization weight, i.e. $\lambda=0.75$)~\cite{tourbier_mialsrtk_2020}.
Volumes are aligned into a common reference space and resampled to $1.1 mm^3$.
Following the FeTA annotation guidelines~\cite{payette_automatic_2021}, the extra-axial cerebrospinal fluid spaces (CSF), the cortical gray matter (cGM), the white matter (WM), the ventricular system% (lateral, third and fourth ventricles)
, the cerebellum, the deep gray matter (dGM) and the brainstem were manually annotated.

\subsection{Out-of-SR domain pure testing set -- \textit{FeTA-KCL}}
\vspace{-.2cm}
Fourty (40) clinical fetal brain images from FeTA dataset are used as out-of-domain pure testing set~\cite{payette_automatic_2021, fidon_labelset_2021}. Subjects were aged from 21.2 to 34.8 weeks of GA ($27.1 \pm 3.9$). They were reconstructed with SIMPLE-IRTK\footnote{Simplified version of the Image Registration Toolkit, Ixico Ltd. licence}~\cite{kuklisova-murgasova_reconstruction_2012,payette_automatic_2021}. 
SR volumes were resampled to an isotropic resolution of $0.8 mm$. Tissue annotations of the CSF, the cGM, the WM, the ventricles, the cerebellum, the dGM and the brainstem were manually refined and completed with the additional corpus callosum (CC) label~\cite{fidon_labelset_2021}.
For the remainder of this study, CC and WM tissue classes are merged in order to match the tissue distribution available in \textit{CHUV-set} (see Section~\ref{sssec:feta_chuv}).

\section{Methodology}
\label{sec:methodology}
\vspace{-.2cm}
Our experiment design is shown in Fig.~\ref{fig:experiment_design} including datasets (A and B), the methods development (C) and their assessment (D).
We propose a single-pipeline multi-reconstruction approach as data augmentation for fetal brain MRI segmentation. 
To strengthen the generalization of our findings, we assess our multi-reconstruction approach with two different SR pipeline, namely NiftyMIC~\cite{ebner_automated_2020} and MIALSRTK~\cite{tourbier_mialsrtk_2020}.
First, we assess our single-pipeline multi-reconstruction method in a pure data augmentation set up (Task 1, Section~\ref{ssec:evaluation}). Second, we further evaluate our augmentation approach in an out-of-domain experiment (Task 2, Section~\ref{ssec:evaluation}).

\subsection{SR reconstruction-based data augmentation}

\vspace{-.2cm}
\textbf{SR reconstructions.} All clinical MR exams in Section~\ref{ssec:clinical_exams} are additionally reconstructed through two independent SR pipelines, such that we have the following two new independent HR \textit{Multi-SR} datasets:
\begin{description}
    \item[\textit{Multi-MIALSRTK}] subjects are reconstructed through the MIALSRTK~\cite{tourbier_mialsrtk_2020} pipeline with the following regularization weight %-  including the default one - 
    $\lambda \in \{0.1,0.75,1.5,3.0\}$;
    \item[\textit{Multi-NiftyMIC}] subjects are reconstructed through the Nifty-MIC \cite{ebner_automated_2020}  pipeline with the regularization weight $\alpha \in \{0.01,\allowbreak 0.02,\allowbreak 0.05,\allowbreak 0.1\}$.
\end{description}

\begin{figure}[h!]
    \centering
    \includegraphics[width=0.86\linewidth]{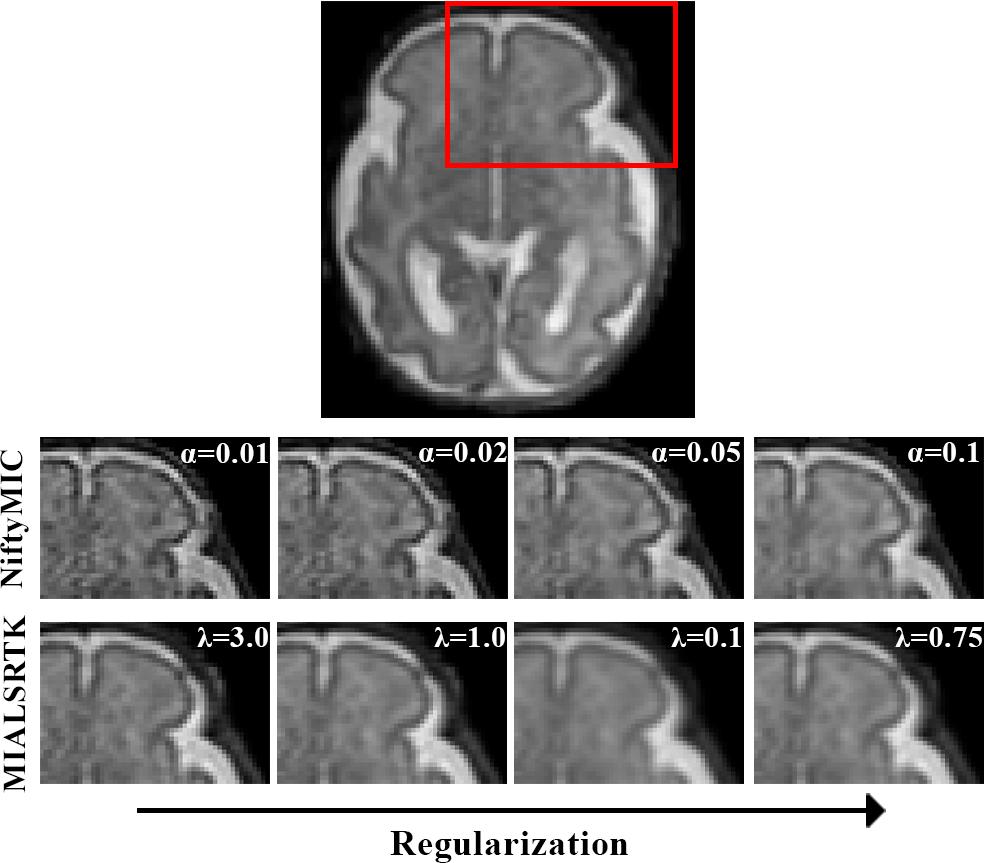}
    \caption{Illustration of the intensity variation depending on the regularization parameters $\lambda$ and $\alpha$ for MIALSRTK and NiftyMIC reconstructions on a 32 weeks of GA %neurotypical 
    subject.}
\label{fig:results_regularization}
\end{figure}

\noindent Fig.~\ref{fig:results_regularization} shows a representative patch of a 32 weeks of GA fetal brain multi-reconstructed through MIALSRTK and NiftyMIC SR pipelines. 
High $\lambda$ (respectively low $\alpha$), offers a better tissue contrast, although the overall image appears more noisy. Conversely, low $\lambda$ values (respectively high $\alpha$) increase the overall smoothness of the image. Thus, variation of the regularization weights echoes a \emph{texture} change in the SR image. 

% \vspace{0.2cm}
\noindent\textbf{Weak labelling.} All reconstructed brain of \textit{Multi-MIALSRTK} and \textit{Multi-NiftyMIC} has a tissue labelmap in the space of its \textit{CHUV-set} reconstruction.
Through a rigid registration-based approach, manual annotations are propogated to the newly generated \textit{Multi-SR} sets.

\vspace{0.2cm}
\noindent\textbf{Training configurations.}
We define three configurations based on their trainining data (Table~\ref{tab:training_configuration_summary}, Fig.~\ref{fig:experiment_design}(C)). \textit{Baseline} is trained on the 30 training subjects from \textit{CHUV-set}, and \textit{MIALSRTK-augmented}, resp. \textit{NiftyMIC-augmented} are trained on the 120 \textit{Multi-MIALSRTK}, resp. \textit{Multi-NiftyMIC}, reconstructed volumes of the same 30 fetal brains.

\begin{table}[h!]
    \centering
\large
    \resizebox{\linewidth}{!}{
\begin{tabular}{l ccc }
\toprule
    & \textit{Baseline} & \textit{MIALSRTK-augmented} & \textit{NiftyMIC-augmented} \\
\midrule
Training set       & \textit{CHUV-set} & \textit{Multi-MIALSRTK} & \textit{Multi-NiftyMIC} \\
$\#$ subjects      & 30 & 30 & 30 \\
$\#$ volumes    & 30 & 120 & 120 \\
Labelling           & Manual & Weak & Weak \\
\bottomrule
\end{tabular} }
\caption{Summary of training configurations.}
\label{tab:training_configuration_summary}
\end{table}

\subsection{Evaluation}
\label{ssec:evaluation}
\vspace{-.2cm}
We compare our multi-reconstruction approach in:
\begin{itemize}
    \item[(a)] Data augmentation task. In this in-domain experiment, we compare the performances of \textit{Baseline} and \textit{MIALSRTK-augmented} on the 10 left-out subject from \textit{CHUV-set}.
    \item[(b)] Domain generalization task. We evaluate the performance of \textit{Baseline}, \textit{MIALSRTK-augmented} and \textit{NiftyMIC-augmented} on the 40 out-of-domain \textit{FeTA-KCL} images. 
\end{itemize}
The performance of the \textit{SR-augmented} and \textit{Baseline} configurations are evaluated with the Dice similarity coefficient (DSC)~\cite{dice_1945} and the average symmetric surface distance (ASSD)~\cite{yeghiazaryan_family_2018} between the ground truth manual annotations and the predicted segmentation. 
A paired Wilcoxon rank-sum test is performed between \textit{SR-augmented} configuration and the \textit{Baseline}. \textit{p}-values are adjusted for multiple comparisons using Bonferroni correction in the statistical analysis of individual fetal brain tissues. Statistical significance level is set to 0.05.

\subsection{Model and training strategy}
\label{ssec:model_training}

From the MONAI framework~\cite{monai_consortium_monai_2022}, we use the popular 3D U-Net architecture that performed remarkably on fetal brain MRI tissue segmentation in the 2021 MICCAI FeTA challenge~\cite{payette_fetal_2022}. Inputs are $96\times96\times96$ voxel size patches that are randomly sampled from the original SR volumes.
At train time, input samples are augmented, based on the most applied and successful transformations used in the FeTA challenge ~\cite{payette_fetal_2022}. Spatial (flipping, rotation, resampling) and intensity-based (bias field, gaussian noise) transformations are randomly applied. Lastly, image patch intensities are normalized.
We adopt a 5-folds cross-validation (CV). Networks are trained for $100$ epochs minimizing a dice focal loss function, where both losses equally contribute.
All configuration adopt the same training strategy.

At test-time, we proceed to an ensemble evaluation of all 5 CV networks on 50\% overlapping patches inferred through a sliding window approach.

\vspace{-.2cm}
\section{Results}
\label{sec:Results}
\vspace{-.2cm}

\begin{table*}[h!]
    \centering
    \resizebox{0.76\linewidth}{!}{
\begin{tabular}{ll cc ccc }
\toprule
    & & \multicolumn{2}{c }{\textbf{(a) Data augmentation}} & \multicolumn{3}{c}{\textbf{(b) Domain generalization}}  \\
    % \cmidrule(l){2-3} \cmidrule(l){4-6}
    \cmidrule(l){3-4} \cmidrule(l){5-7}
    & & \textit{Baseline} & \textit{MIALSRTK-augmented} & \textit{Baseline} & \textit{MIALSRTK-augmented} & \textit{NiftyMIC-augmented} \\
\midrule
\multirow{8}{*}{DSC ($\uparrow$)} & CSF         &  ~0.87 $\pm$ 0.02~ &  \textbf{0.90 $\pm$ 0.02} (*) & ~0.87 $\pm$ 0.17~    & 0.90 $\pm$ 0.16 (*)   & \textbf{0.90 $\pm$ 0.17} (*)  \\
& cGM         &  ~0.75 $\pm$ 0.03~ &  \textbf{0.78 $\pm$ 0.03} (*) & ~0.76 $\pm$ 0.14~    & \textbf{0.81 $\pm$ 0.14} (*)   & 0.81 $\pm$ 0.14 (*)  \\
& WM          &  ~0.86 $\pm$ 0.02~ &  \textbf{0.89 $\pm$ 0.01} (*) & ~0.85 $\pm$ 0.13~    & \textbf{0.89 $\pm$ 0.12} (*)   & 0.88 $\pm$ 0.12 (*)  \\
& Ventricles  &  ~0.81 $\pm$ 0.05~ &  \textbf{0.84 $\pm$ 0.04} (*) & ~0.76 $\pm$ 0.11~    & 0.82 $\pm$ 0.10 (*)   & \textbf{0.86 $\pm$ 0.10} (*)  \\
& Cerebellum  &  ~0.88 $\pm$ 0.02~ &  \textbf{0.91 $\pm$ 0.03} (*) & ~0.67 $\pm$ 0.28~    & 0.79 $\pm$ 0.21 (*)   & \textbf{0.86 $\pm$ 0.11} (*)  \\
& dGM         &  ~0.88 $\pm$ 0.05~ &  \textbf{0.89 $\pm$ 0.03} (*) & ~0.79 $\pm$ 0.11~    & 0.85 $\pm$ 0.08 (*)   & \textbf{0.86 $\pm$ 0.06} (*)  \\
& Brainstem   &  ~0.85 $\pm$ 0.02~ &  \textbf{0.87 $\pm$ 0.02} (*) & ~0.68 $\pm$ 0.20~    & 0.76 $\pm$ 0.13 (*)   & \textbf{0.76 $\pm$ 0.10}      \\
% \midrule
\cmidrule(l){2-7}
& Overall   &   0.84 $\pm$ 0.02 & \textbf{0.87 $\pm$ 0.01} (*) & 0.77 $\pm$ 0.14 & 0.83 $\pm$ 0.11 (*) & \textbf{0.85 $\pm$ 0.10} (*) \\
\midrule
ASSD ($\downarrow$) & Overall   &   0.70 $\pm$ 0.07 & \textbf{0.56 $\pm$ 0.05} (*) & 1.64 $\pm$ 1.32 & 1.14 $\pm$ 0.94 (*) & \textbf{1.03 $\pm$ 0.80} (*) \\
\bottomrule
\end{tabular} }
\caption{DSC and ASSD (mean $\pm$ SD) of the different training configurations in data augmentation (a) and domain generalization (b) tasks. The best scores between \textit{SR-augmented} configurations and \textit{Baseline} are shown in bold. Arrows indicates weither the metric is better maximized ($\uparrow$) or minimized ($\downarrow$). The corresponding \emph{p}-values (paired Wilcoxon rank sum test) are adjusted for multiple comparisons using Bonferroni correction. Statistical significance (*) is $p<0.05$.}
\label{tab:quantitative_results}
\end{table*}

\subsection{Data augmentation}

Table~\ref{tab:quantitative_results}~(a) reports the DSC and ASSD performances for both task. Overall, the performance of the segmentation algorithm is significantly enhanced when each fetal brain is multi-reconstructed through the same pipeline as target (testing) images, even though the weak labelling process incurred. The benefit of \textit{MIALSRTK-augmentated} is further statistically significant for all tissue classes. 

\subsection{Domain generalization}

\begin{figure*}[h!]
    \centering
    \includegraphics[width=0.83\linewidth]{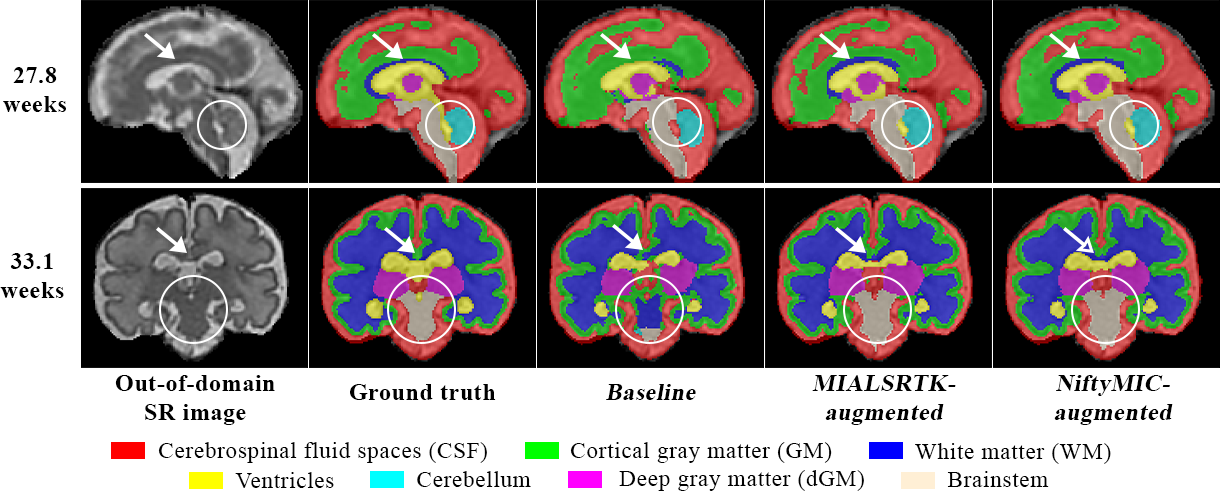}
    \caption{Sagittal view of a 27.8 weeks-old (top) and coronal view of the 33.1 weeks-old (bottom) fetal brain tissue segmentation obtained in the different configurations studied. White arrows and circles show representative areas where our multi-reconstruction approach improves the segmentation accuracy.}
\label{fig:results_qualitative}
\end{figure*}

\begin{figure}[h!]
    \centering
    \includegraphics[width=\linewidth]{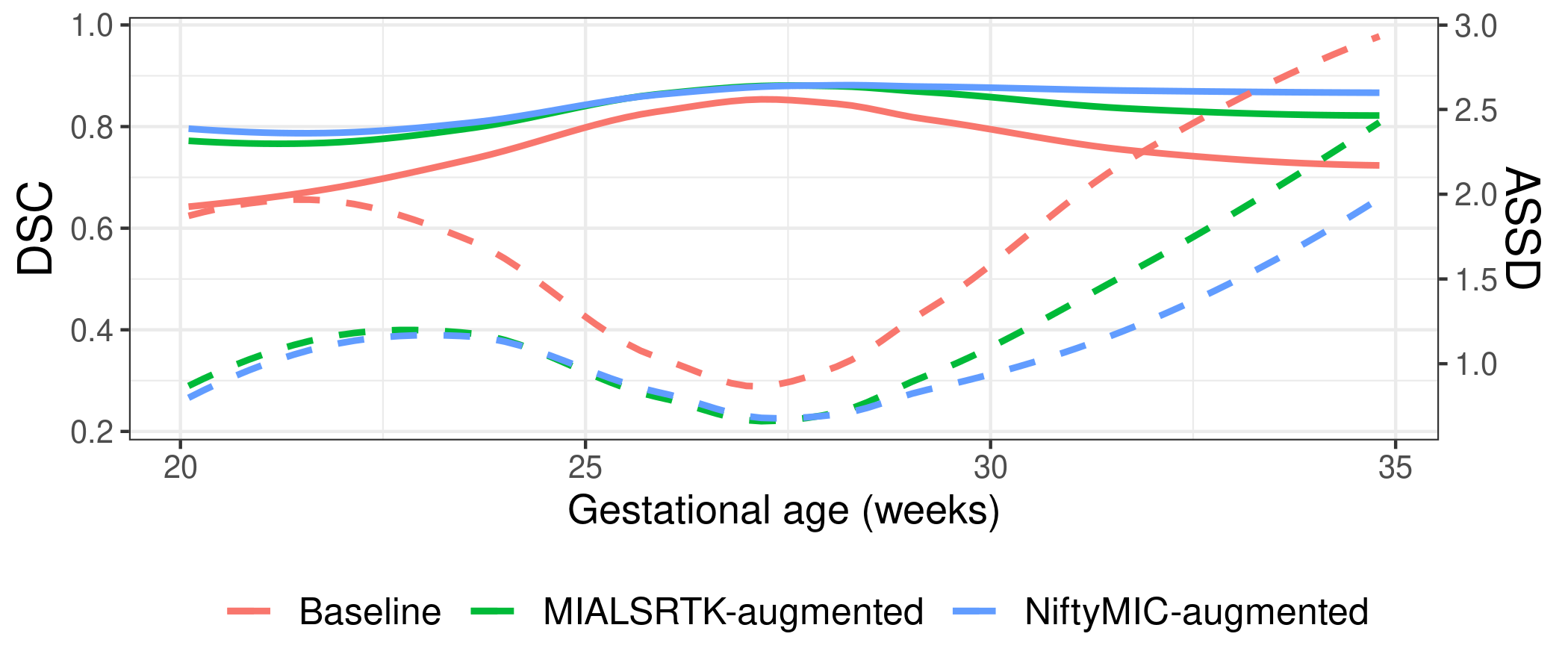}
    \caption{Mean DSC (plain) and ASSD (dashed)  performance in the domain generalization task as a function of GA in weeks.}
    \label{fig:results_task2_f_ga}
\end{figure}
\textit{SR-augmented} qualitatively enhance the accuracy of segmentation compared to \textit{Baseline}, especially in the infratentorial structures, namely the brainstem, the cerebellum and the $4^{th}$ ventricle (Fig.~\ref{fig:results_qualitative}, white circles).
Additionally, the WM tract of the CC are better captured with our multi-reconstruction approach, and even more than compared to the ground truth.

Quantitative results (Table~\ref{tab:quantitative_results} (b)) show that performance from 10 in-domain subjects to 40 out-of-domain subjects causes a loss in the performances of both \textit{Baseline} 
%(from $0.84$ (a) to $0.77$ (b)) 
and \textit{MIALSRTK-augmented} 
%(from $0.87$ (a) to $0.83$ (b)) 
configurations. % Nevertheless, while \textit{Baseline} shows a loss in DSC of $0.07$, \textit{MIALSRTK} only drops of $0.04$. 
Nevertheless, while \textit{Baseline} shows a loss of $0.07$ and $0.94$, respectively in DSC and ASSD \textit{MIALSRTK-augmented} only drops of $0.04$ and $0.58$.
Our multi-reconstruction approach hence seems more robust to the inter-SR method domain gap.
Regardless of the SR method of the training data, our multi-reconstruction augmentation strategy significantly improves the segmentation performance on out-of-SRR domain images, both in DSC and ASSD. 
On a tissue-wise analysis, \textit{SR-augmented} configurations are always significantly better performing than \textit{Baseline}. The benefit of our multi-reconstruction approach is even more pronounced in the ventricles, the cerebellum, the dGM and the brainstem where the gain in DSC is greater than $0.05$. Specifically, the steepest improvement appears in the cerebellum ($0.67$ for the \textit{Baseline} vs. $0.79$ and $0.86$, resp. for \textit{MIALSRTK-} and \textit{NiftyMIC-augmented}).
From a GA-based analysis (Fig.~\ref{fig:results_task2_f_ga}), we observe that \textit{Baseline} overall performs worse on young and old ($<25$ and $>30$ weeks of GA) fetuses. On the contrary, although a similar trend is noticeable for \textit{SR-augmented} methods, it is substantially less pronounced. 
Consequently, in this inter-domain segmentation task, \textit{SR-augmented} configurations seems to better benefit the endpoint of the GA range studied.

\vspace{-.2cm}
\section{Conclusion}
\vspace{-.2cm} 
We have demonstrated that having single-pipeline multi-reconstruction of fetal brain MR exams \textit{(i)} is an efficient intensity-based data augmentation strategy and \textit{(ii)} reduces the performance drop in target image domain shift in segmentation task.
Our multi-reconstruction approach, combined to conventional data augmentation strategies, increases the representation of fetal brain MRI variability in the training phase of supervised segmentation method. 
Although we did not investigate multi-pipeline multi-reconstruction augmentation, one can expect an even stronger benefit of our method.
In its batch processing approach, our multi-reconstruction strategy is an out-of-the-box easy to adapt method. 
% Future work will investigate the logical step of aggregating multi-pipeline multi-reconstruction augmentation. 
Future work will investigate this multi-reconstruction augmentation at inference in order to increase the prediction robustness.

% \clearpage
% \section{Statement of novelty (296 / 300 characters)}
% \textcolor{red}{Our work presents a novel single-method multi-parameter SR reconstruction approach of fetal brain MRI to overcome the lack of annotated data. We validate our method on two tasks: data augmentation and domain generalization of fetal brain tissue segmentation. }

\clearpage

\section{Compliance with ethical standards}
\label{sec:ethics}
The local ethics committee of the Canton of Vaud, Switzerland (CER-VD 2021-00124) 
approved the retrospective collection and analysis of MRI data and the prospective studies for the collection and analysis of the MRI data in presence of a signed form of either general or specific consent.

\section{Acknowledgments}
\label{sec:acknowledgments}

This work is supported by the Swiss National Science Foundation through grants 182602 and 141283. We acknowledge access to the facilities and expertise of the CIBM Center for Biomedical Imaging, a Swiss research center of excellence founded and supported by Lausanne University Hospital (CHUV), University of Lausanne (UNIL), Ecole polytechnique fédérale de Lausanne (EPFL), University of Geneva (UNIGE) and Geneva University Hospitals (HUG).
The authors have no relevant financial or non-financial interests to disclose.

\text

\bibliographystyle{IEEEbib}
\bibliography{strings,refs}

\end{document}